\documentclass[conference]{IEEEtran}
\IEEEoverridecommandlockouts

\usepackage[english]{babel}
\usepackage[T1]{fontenc}
\usepackage{amssymb, amsmath}
\usepackage{cite}
\usepackage{graphicx}
\usepackage{subfigure}
\usepackage{xcolor}
\usepackage{hyperref}

\newcommand{\norm}[2]{\| #1 \|_{#2}}

\newcommand{\cscgdist}[2]{\sim \mathcal{CN} \left( #1, #2 \right)}
\newcommand{\complexset}{\mathbb{C}}
\newcommand{\realset}{\mathbb{R}}
\newcommand{\integerset}{\mathbb{Z}}
\newcommand{\set}[1]{\left\{ #1 \right\}}

\def\BibTeX{{\rm B\kern-.05em{\sc i\kern-.025em b}\kern-.08em T\kern-.1667em\lower.7ex\hbox{E}\kern-.125emX}}

\begin{document}

% Title
\title{Energy Efficiency and Throughput of Random Access Protocols for RIS-Aided IoT Networks}

% Authors and funding information
\author{
\IEEEauthorblockN{Jo\~ao Henrique Inacio de Souza}
\IEEEauthorblockA{\textit{Electrical Eng. Dept.}\\
\textit{UEL}\\
Londrina, Brazil\\
joaohis@outlook.com}

\and

\IEEEauthorblockN{Jos\'e Carlos Marinello Filho}
\IEEEauthorblockA{\textit{Electrical Eng. Dept.}\\
\textit{UTFPR}\\
Corn\'elio Proc\'opio, Brazil\\
jcmarinello@utfpr.edu.br}

\and

\IEEEauthorblockN{Taufik Abr\~ao}
\IEEEauthorblockA{\textit{Electrical Eng. Dept.}\\
\textit{UEL}\\
Londrina, Brazil\\
taufik@uel.br}

\and

\IEEEauthorblockN{Cristiano Panazio}
\IEEEauthorblockA{\textit{Telecommunications  Dept.}\\
\textit{EPUSP}\\
S\~ao Paulo, Brazil\\
cpanazio@usp.br}

\thanks{This work was supported by the National Council for Scientific and Technological Development (CNPq) of Brazil under Grants 405301/2021-9, 141445/2020-3, and 310681/2019-7.}
}

\maketitle

% Abstract
\begin{abstract}
Green Internet of Things (IoT) aims to enable a sustainable smart world by making energy efficiency (EE) the main performance indicator for IoT hardware and software.
With respect to network design, this implies in developing energy-efficient communication protocols and network architectures adapted to the ubiquitousness of the IoT machine-type devices (MTDs) and the sporadic traffic generated by them, keeping a low power consumption at the MTDs-side.
In this sense, reconfigurable intelligent surfaces (RISs) have presented the capacity of significantly improving the network coverage using mostly passive reflecting elements, drastically reducing the power expenditure.
In this paper, we develop a realistic power consumption model and an expression for the overall system EE for RIS-aided IoT networks that adopt a two time-scale random access (RA) protocol to handle the uplink transmissions.
Specifically, during each time slot of the RA protocol, the RIS covers a specific area of interest in the communication cell with a predefined set of phase-shift configurations, changing the channel qualities of the contending MTDs.
Numerical results comparing the RA protocol performance reveal that access policies that exploit information of the channel qualities are suitable for green IoT networks, simultaneously attaining competitive EE and throughput combined with low power consumption at the MTDs-side.
\end{abstract}

% Keywords

% Introduction
\section{Introduction}\label{sec:introduction}

The concept of green IoT brings the challenge of developing a sustainable IoT environment by paths such as designing energy-efficient systems and devices, and prioritizing the use of renewable power sources \cite{zhu2015}.
From the communication protocol standpoint, this objective leads to optimized schemes that make an efficient use of radio resources, including  transmit power and time-frequency resource blocks.
In this sense, \cite{chen2022} has recently proposed a framework to improve the devices’ EE by optimizing the transmit power and the maximum number of packet replicas in the \textit{irregular repetition slotted} ALOHA (IRSA) protocol.
Although optimized protocols are able to offer increased EE, combining these designs with energy-efficient technologies can further enhance these gains.

RIS is a promising low-cost and low-power technology to enable sustainable wireless networks.
This technology can produce anomalous reflection of the incoming signal to create hot spots in areas of interest in the communication cell without any power amplifier (PA), consuming much less energy than common active transceiver technologies, such as arrays with massive numbers of antennas.
Specifically, in the analysis of \cite{huang2019}, it is shown that RIS-aided networks can achieve up to a threefold increase in EE when compared with networks assisted by multiple-antenna amplify-and-forward relays.

Recently, \cite{croisfelt2022} has proposed a two time-scale RA protocol for RIS-aided networks in which the RIS has a limited set of phase-shift configurations, designed to cover specific areas of interest in the communication cell. During the first phase of the protocol, the contending MTDs measure their channel qualities over a sequence of training time slots, where in each slot the RIS induces a different phase-shift configuration. Then, at the second phase, the RIS repeats the configuration change pattern over a sequence of access time slots, where in each slot the MTDs independently decide to transmit according to the measured channel qualities.
This protocol increases the average number of \textit{successful access attempts} at the expense of increased access delay, which is a minor issue in delay-tolerant IoT applications.

In this paper, we develop a realistic power consumption model and the throughput formula for RIS-aided IoT networks that adopt the RA protocol proposed in \cite{croisfelt2022}. With this model, we derive an expression for the overall system EE, evaluating the protocol performance with four different access policies. Moreover, we extend the \textit{strongest-configuration access policy} (SCP) given in \cite{croisfelt2022} to fully exploit the successive interference cancellation (SIC) receiver implemented at the access point (AP)-side. Finally, we adapt the \textit{contention resolution diversity slotted ALOHA} (CRDSA) \cite{casini2007} and the state-of-the-art IRSA \cite{liva2011} protocols to create two novel access policies suitable for IoT networks. 

% System model
\section{System Model}\label{sec:system-model}

\begin{figure}[t]
\centering
\subfigure[TDMA frame]{
\includegraphics[width=\columnwidth]{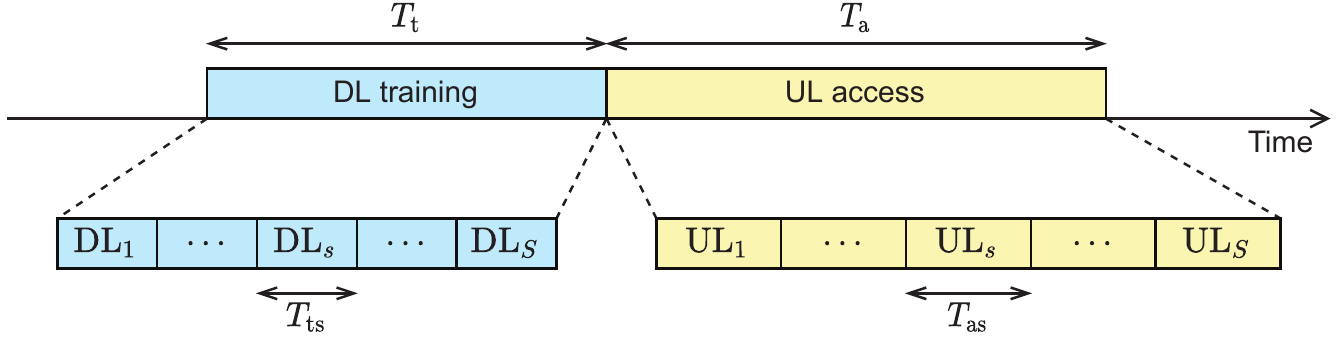}
\label{fig:tdma-frame}
}
\subfigure[RIS and communication cell setup]{
\includegraphics[width=\columnwidth]{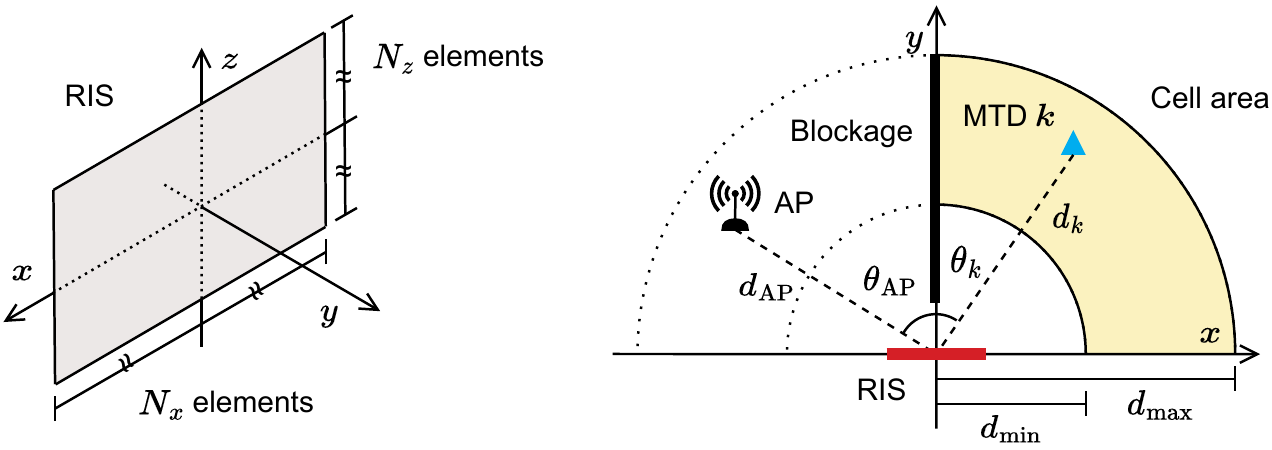}
\label{fig:ris-communication-cell-setup}
}
\caption{a) Diagram of the TDMA frame, and b) setup of the RIS-aided communication scheme.}
\label{fig:ris-commmunication-cell-tdma-frame}
\end{figure}

In this section, we describe the time-division multiple access (TDMA) frame, the RIS-aided communication scheme, and the two time-scale RA protocol.
The TDMA frame, depicted in Fig. \ref{fig:tdma-frame}, is comprised by a downlink (DL) training block of duration $T_\text{t} > 0$ and an uplink (UL) access block of duration $T_\text{a} > 0$. The training and access blocks are divided into $S \in \integerset_+$ time slots with lengths respectively equal to $T_\text{ts} > 0$ and $T_\text{as} > 0$, \textit{i.e.}, $T_\text{t} = ST_\text{ts}$ and $T_\text{a} = ST_\text{as}$. The two blocks of the TDMA frame are used by the RA protocol to handle the UL transmissions of the contending MTDs, namely the MTDs that want to transmit a data burst during an RA frame.

Fig. \ref{fig:ris-communication-cell-setup} depicts the RIS and communication cell setup.
The RIS is a thin surface located at the $xz$-plane with center at the origin. The surface is constituted by $N_x \in \integerset_+$ reflecting elements in the $x$-axis direction and $N_z \in \integerset_+$ in the $z$-axis direction, totaling $N = N_x N_z$ elements. Each reflecting element has a rectangular surface of sides $d_x,d_z \leq \lambda$, where $\lambda$ denotes the carrier wavelength.
Similarly to \cite{croisfelt2022}, we assume that the signal reflection produced by the RIS is independent of the $z$-dimension.
We consider that the RIS has a fixed number of phase-shift configurations equal to the number of time slots, $S$. Specifically, the set of phase-shift configurations of the RIS as a function of the number of time slots is denoted by:
\begin{equation}
\Theta(S) = \set{\left. \frac{\pi(s-1)}{2(S-1)} \right\vert s \in \set{1,\dots,S}}.
\end{equation}
Therefore, the $s$-th phase-shift configuration induced by the RIS in the impinging signal is denoted by $\theta_s \in \Theta(S)$. Since in the RA protocol we assume that the RIS has a dedicated phase-shift configuration for each time slot, throughout the text we use $s$ to denote the $s$-th RIS configuration and the $s$-th time slot interchangeably.

In the proposed setup, the AP and MTDs are in the $xy$-plane, located respectively in the fourth and first quadrants.
Specifically, the AP is located at a distance $d_{\textsc{ap}} \in [d_{\min}, d_{\max}]$ from the origin, with $0 < d_{\min} < d_{\max}$, forming an angle $\theta_{\textsc{ap}} \in [0, \pi/2]$ with the RIS boresight. Similarly, the MTD $k$ is located at a distance $d_k \in [d_{\min}, d_{\max}]$ from the origin, forming an angle $\theta_k \in [0, \pi/2]$ with the RIS boresight.
The contending MTDs communicate with the AP via a virtual line-of-sight path created by the RIS, since the first and fourth quadrants are separated by a blockage that prevents direct radio links between them.

Let $\alpha_{\textsc{ap}} > 0$ and $\alpha_k > 0$ be the AP and the $k$-th MTD antenna gains, respectively. The channel coefficient of the MTD $k$ during the access time slot $s$, denoted by $h_k(s) \in \complexset$, is equal to \cite{ozdogan2020}:
\begin{equation}
h_k(s) = \sqrt{\beta_k} e^{j\psi_k} \Omega_k(s),
\end{equation}
where $\beta_k \in \realset$ is the path loss, $\psi_k \in \realset$ is the total phase-shift, and $\Omega_k(s) \in \complexset$ is the RIS array factor \cite{croisfelt2022}, respectively defined by:
\begin{gather}
\beta_k = \frac{\alpha_{\textsc{ap}} \alpha_k}{(4\pi)^2} \left(\frac{d_x d_z}{d_{\textsc{ap}} d_k}\right)^2 \cos^2 \theta_k,\\
\psi_k = \omega \left[ d_{\textsc{ap}} + d_k - (\sin \theta_{\textsc{ap}} - \sin \theta_k) \frac{N_x + 1}{2} d_x\right],\\
\Omega_k(s) = N_z \sum_{n = 1}^{N_x} e^{j\omega (\sin \theta_k - \sin \theta_s) n d_x},
\end{gather}
where $\omega = 2\pi/\lambda$ denotes the wavenumber.

Let $\mathcal{K}$ be the set with the indices of the contending MTDs during an RA frame, and such that its cardinality is $|\mathcal{K}| = K$ with $K \in \integerset_+$. In addition, let $\mathcal{K}(s) \subseteq \mathcal{K}$ be the set with the indices of the contending MTDs that transmit during the access time slot $s$. Therefore, the UL received signal at the AP during the access time slot $s$ is equal to:
\begin{equation}
\label{eq:uplink-signal}
\mathbf{y}(s) = \sum_{k \in \mathcal{K}(s)} \sqrt{\rho_{\textsc{mtd}}} h_k(s) \mathbf{v}_k + \mathbf{w}(s),
\end{equation}
where $\rho_{\textsc{mtd}}$ is the MTD transmit power, $\mathbf{v}_k \in \complexset^L$ such that $\norm{\mathbf{v}_k}{2}^2 = 1$ is the data burst of the MTD $k$ with $L \in \integerset_+$ symbols, and $\mathbf{w}(s) \cscgdist{\mathbf{0}_L}{\sigma_w^2\mathbf{I}_L}$ is the additive noise. From eq. \eqref{eq:uplink-signal}, one can derive the signal-to-noise ratio (SNR) of the MTD $k$ during the access time slot $s$ as:
\begin{equation}
\gamma_k(s) = \frac{\rho_\textsc{mtd} |h_k(s)|^2}{\sigma_w^2}.
\end{equation}
During the DL training, the MTDs measure their respective channel qualities within each time slot by indirectly estimating their SNRs, $\set{\gamma_k(s) \mid s \in \set{1,\dots,S}}$ for each $k \in \mathcal{K}$. For the sake of simplicity but without loss of generality, we consider that the channel qualities measured by the MTDs can be defined as:
\begin{equation}
\label{eq:channel-quality}
q_k(s) = c_k \gamma_k(s) + \varepsilon_k(s)
\end{equation}
for $k \in \mathcal{K}$, and $s \in \set{1,\dots,S}$, where $c_k \in \realset$ is a constant to account for the DL channel quality, and $\varepsilon_k(s) \in \realset$ is the estimation error such that $\mathbb{E}[\varepsilon_k(s)] = 0$ and $\mathbb{V}\mathrm{ar}[\varepsilon_k(s)] = \sigma_{\varepsilon_k}^2$. Further details on the measurement of the channel qualities during the DL training can be found in \cite{croisfelt2022}.

\subsection{Random Access Protocol}\label{sec:random-access-protocol}

In this subsection, we describe the \textit{two phases} that constitute the RA protocol, the DL training phase and the UL access phase.

\begin{figure}[b]
\centering
\vspace{-3mm}
\includegraphics[width=.75\columnwidth]{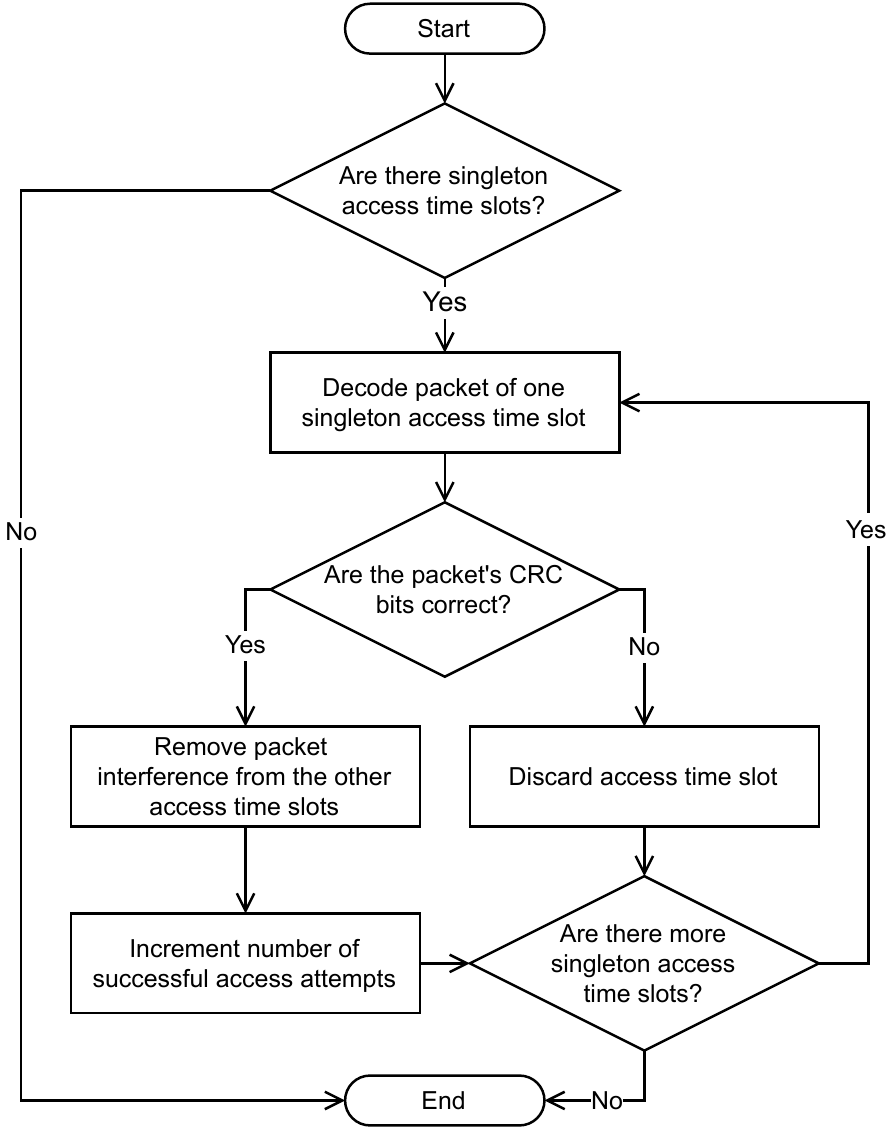}
\caption{Flowchart of the SIC receiver deployed in the AP to decode the signals transmitted by the contending MTDs during the access phase.}
\label{fig:sic-receiver}
\end{figure}

\noindent\underline{Phase \textbf{I}}: Downlink training
\begin{enumerate}

\item AP transmits a training pilot over $S$ consecutive time slots, varying the RIS phase-shift configuration to one of the $S$ available ones during each slot.

\item The contending MTDs listen to the channel during each time slot, estimating and buffering their instantaneous channel qualities as defined in eq. \eqref{eq:channel-quality} for each RIS phase-shift configuration.

\end{enumerate}

\noindent\underline{Phase \textbf{II}}: Uplink access
\begin{enumerate}

\item AP listens to the channel over $S$ consecutive time slots, using the same change pattern of the RIS phase-shift configuration adopted in the training phase.

\item The contending MTDs send packet replicas during a selected set of time slots, chosen independently by each MTD following a predefined access policy. Specifically, four different policies, described in Subsection \ref{sec:access-policies}, are evaluated in this work. Notice that the packet replicas have cyclic redundancy check (CRC) bits appended to enable detection of decoding errors.

\item AP receiver starts by decoding the \textit{singleton time slots}, namely the time slots in which a single packet was transmitted, or, later in the receiver operation, the time slots that have only one remaining packet after interference cancellation (IC). In practice, preamble recognition can be applied to detect the singleton time slots \cite{casini2007}.
After that, a collision resolution procedure based on SIC is carried out to try to decode the colliding packets. The operation of the AP SIC receiver is illustrated in Fig. \ref{fig:sic-receiver}. 
One should note that a packet is successfully decoded if the CRC bits are correct. In this sense, it is considered that the AP receiver can decode a packet if it is within a singleton time slot and its SNR is above the threshold $\gamma_{\text{th}} \in \realset$ in dB. Therefore, the collision resolution between $n$ colliding packets is possible when $n-1$ of those packets are decoded in prior iterations.

\end{enumerate}

% Access Policies
\subsection{Access Policies}\label{sec:access-policies}

Four different access policies have been deployed in the UL access phase of the analyzed RA protocol. Specifically, the \textit{access policy} determines the \textit{access time slots} in which a contending MTD will transmit. Hence, carefully designing the access policies is crucial to enhance the protocol performance by jointly exploiting the spatial diversity among the contending MTDs and the time diversity created by switching between the different RIS configurations.

\vspace{2mm}

\noindent\textbf{1. Configuration-Aware Random Access Policy (CARP).} In CARP \cite{croisfelt2022}, a contending MTD $k$ transmits during each access time slot $s$ with the probability:
\begin{equation}
\label{eq:carp-slot-transmission-probability}
p_k(s) = \frac{q_k(s)}{\sum_{s=1}^S q_k(s)},
\end{equation}
where the channel qualities $q_k(s)$ for $k \in \mathcal{K}$ and $s \in \set{1,\dots,S}$ are defined in \eqref{eq:channel-quality}. With these probabilities, the contending MTDs are more likely to transmit within the access time slots that provided the best channel qualities during the DL training phase. In CARP, the decision of transmitting during a time slot is made from Bernoulli trials following the probabilities defined by \eqref{eq:carp-slot-transmission-probability}. If no time slot is selected, the contending MTD will transmit during the time slot that provided the best channel quality. Therefore, each contending MTD can transmit a single packet or up to $S$ packet replicas.

\vspace{2mm}

\noindent\textbf{2. $s$-Strongest-Configurations Access Policy ($s$-SCP).} This policy is a  generalization of the SCP, originally introduced in \cite{croisfelt2022}, to fully exploit the benefits of the SIC procedure at the AP's receiver. In $s$-SCP, a contending MTD selects the $s \in \integerset_+$ access time slots that provided the best channel qualities during the DL training phase. Hence, in this access policy, each contending MTD transmits $s$ packet replicas, totaling $sK$ packet replicas during the UL access phase.
It is straightforward to notice that $s$-SCP is equivalent to the original SCP if $s = 1$, and SIC cannot be applied. Conversely, SIC can be conveniently applied for $s \geq 2$.

\vspace{2mm}

\noindent\textbf{3. Contention Resolution Diversity Slotted ALOHA Access Policy (CRDSAP).} This policy is an adaption of the CRDSA protocol originally proposed in \cite{casini2007} as an extension of the \textit{slotted ALOHA} with support to IC techniques. In CRDSAP, a contending MTD transmits during $s = 2$ access time slots chosen randomly with the same probability, totaling $2K$ transmitted packet replicas during the UL access phase. As the access time slots are chosen randomly without any information about the channel qualities, CRDSAP does not require the DL training phase of the RA protocol. The essential difference between the CRDSAP and the CRDSA protocol is that, during each time slot of the former, the contending MTDs experience different channel conditions due to the changing RIS phase-shift configuration.
This creates spatial and time diversities that, as it will be demonstrated further, can increase the system EE and throughput.

\vspace{2mm}

\noindent\textbf{4. Irregular Repetition Slotted ALOHA Access Policy (IRSAP).} This policy is an adaption of the IRSA protocol proposed in \cite{liva2011}, introducing variable packet repetition rates to increase the throughput attained by CRDSA.
In IRSAP, a contending MTD transmits $s$ packet replicas following a statistical distribution defined by the probability generating function \cite{chen2022}:
\begin{equation}
\label{eq:repetition-rate-distribution}
\Lambda(x) = \left( 1 + \frac{1}{S-1} \right) \sum_{s=2}^{S} \frac{x^s}{(s-1)s}.
\end{equation}
The $s$ access time slots selected to transmit the replicas are chosen with the same probability. Since the access time slots are chosen randomly without any information about the channel qualities, IRSAP does not require the DL training phase.
The difference between IRSAP and the original IRSA protocol is the spatial and time diversities created within the UL access phase by changing the RIS configuration between the time slots.
In this policy, each MTD can transmit 2 up to $S$ packet replicas, resulting in the average total number of transmitted packet replicas:
\begin{equation}
K \left( 1 + \frac{1}{S-1} \right) \sum_{s=2}^{S} \frac{1}{s-1},
\end{equation}
which can be obtained straightforwardly from eq. \eqref{eq:repetition-rate-distribution}.

% System Power Consumption Model
\subsection{System Power Consumption Model}

In this subsection, we present the power consumption model, an extension of the model proposed in \cite{huang2019} adapted to account for the power consumed during the RA protocol phases.

Let $\bar{P}_{\textsc{ap}} > 0$ be the static power dissipated by the AP hardware, $\rho_{\textsc{ap}} > 0$ be the AP transmit power during each training time slot, and $\xi_{\textsc{ap}} > 1$ be the inverse of the AP transmit PA efficiency. The power consumed by the AP is equal to:
\begin{equation}
\label{eq:ap-consumed-power}
P_\textsc{ap} = S\xi_\textsc{ap}\rho_\textsc{ap} + \bar{P}_\textsc{ap}.
\end{equation}
To derive the power consumption of the RIS, we consider that each of the $N$ reflecting elements is constituted by a phase-shifter with limited resolution \cite{huang2019}.
Let $P_n(b) > 0$ be the power consumption of a phase-shifter having $b$-bit resolution. The power consumed by the RIS is equal to:
\begin{equation}
\label{eq:ris-consumed-power}
P_\textsc{ris} = N P_n(b).
\end{equation}
Finally, let $\bar{P}_{\textsc{mtd}} > 0$ be the static power consumed by the hardware of each MTD. In addition, let $s_k \in \set{1,\dots,S}$ be the number of packet replicas sent by the contending MTD $k$ during an RA frame, and $\xi_{\textsc{mtd}} > 1$ be the inverse of the transmit PA efficiency of each MTD. The power consumed by the contending MTD $k$ is equal to:
\begin{equation}
\label{eq:mtd-consumed-power}
P_{\textsc{mtd},k} = s_k\xi_\textsc{mtd}\rho_\textsc{mtd} + \bar{P}_\textsc{mtd}.
\end{equation}
With the power consumed by the AP, RIS, and contending MTDs, we can write the power consumed by the system during an RA frame as:
\begin{equation}
\label{eq:consumed-power}
P = P_\textsc{ap} + P_\textsc{ris} + \sum_{k \in \mathcal{K}} P_{\textsc{mtd},k}.
\end{equation}

% System Energy Efficiency
\section{System Energy Efficiency}\label{sec:system-energy-efficiency}

Considering the RA protocol described in Section \ref{sec:system-model}, a \textit{successful access attempt} (SA) occurs when, during the access phase, one packet replica transmitted by a contending MTD can be successfully decoded at the SIC receiver.
Therefore, let $A \leq K$ be the number of SAs during the RA frame. From Fig. \ref{fig:tdma-frame}, the system throughput can be calculated by:
\begin{equation}
\label{eq:throughput}
G = \frac{A}{T_{\text{t}} + T_{\text{a}}}
= \frac{A}{ST_{\text{ts}} + ST_{\text{as}}}
= \frac{A}{(1+r)ST_{\text{as}}} \text{[packet/s]},
\end{equation}
where, in the last equality, $r =\frac{T_{\text{ts}}}{T_{\text{as}}}$ is the ratio between the duration of the training and the access time slots, namely the \textit{training time slot ratio}. One should notice that, for access policies that do not require the DL training phase, such as the CRDSAP and IRSAP, the throughput can be calculated from \eqref{eq:throughput} by setting $r = 0$.

Considering the system throughput in eq. \eqref{eq:throughput} and the power consumption model in eq. \eqref{eq:consumed-power}, the overall system EE is:
\begin{equation}
\label{eq:energy-efficiency}
\eta_{\textsc{ee}} = \frac{G}{P_\textsc{ap} + P_\textsc{ris} + \sum_{k \in \mathcal{K}} P_{\textsc{mtd},k}} \quad \text{[packet/s/W]},
\end{equation}
where the denominator summands are given respectively by eqs. \eqref{eq:ap-consumed-power}, \eqref{eq:ris-consumed-power}, and \eqref{eq:mtd-consumed-power}.

% Numerical Results
\section{Numerical Results}\label{sec:numerical-results}

\begin{table}[b]
\centering
\caption{Parameter values used to generate the numerical results.}
\label{tab:numerical-results-parameter-values}
\begin{tabular}{p{.34\columnwidth}|p{.58\columnwidth}}
\hline
\textbf{Parameter} & \textbf{Description}\\
\hline
$\xi_{\textsc{ap}} = 1.2$ & Inverse of the efficiency of transmit PA\\
$\rho_{\textsc{ap}} = 100$ mW & Transmit power\\
$\bar{P}_{\textsc{ap}} = 9$ dBW & Static power dissipated by the hardware\\
$(d_{\textsc{ap}}, \theta_{\textsc{ap}}) = (20 \text{m}, \pi/4)$ & AP position\\
$\alpha_{\textsc{ap}} = 5$ dB & Antenna gain\\
\hline
$\xi_{\textsc{mtd}} = 1.2$ & Inverse of the efficiency of transmit PA\\
$\rho_{\textsc{mtd}} {\in [1,100]}$ mW & Transmit power\\
$\bar{P}_{\textsc{mtd}} = 40$ mW & Static power dissipated by the hardware\\
$d_k \in [25, 100]$ m & Distance from the origin\\
$\theta_k \in [0, \pi/2]$ & Angle with the RIS boresight\\
$\alpha_k = 5$ dB & MTD antenna gain\\
\hline
${N \in [4,400]}$ ($N_x = N_z$) & Number of RIS elements\\
$P_n(3) = 1.5$ mW & Power consumption of a phase-shifter having 3-bit resolution\\
$d_x = d_z = 10$ cm & RIS elements size\\
\hline
$S \in {[1,40]}$ & Number of time slots\\
$K \in {[2,20]}$ & Number of contending MTDs\\
$T_{\text{as}} = 1$ s & Access time slot duration\\
$r = 0.2$ & Training time slot ratio\\
{$\gamma_{\text{th}} = 0$ dB} & {SNR threshold of the SIC receiver}\\
$\lambda = 10$ cm & Carrier wavelength\\
$\sigma_w^2 = -94$ dBm & Noise power\\
$s = 2$ & $s$-SCP parameter\\
\hline
\end{tabular}
\end{table}

To evaluate the applicability of the RA protocol proposed in \cite{croisfelt2022} to RIS-aided IoT networks, we present numerical results on the \textit{throughput} ($G$) and \textit{EE} ($\eta_{\textsc{ee}}$) considering the four access policies described in Subsection \ref{sec:access-policies}.
The parameter values used to carry out the simulations and calculate the results are given in Table \ref{tab:numerical-results-parameter-values}.
In addition, for simplicity, we assume perfect estimation of the contending MTDs channel qualities, \textit{i.e.}, in \eqref{eq:channel-quality}, $c_k = 1$ and $\varepsilon_k(s) = 0$ for all $k \in \mathcal{K}$ and $s \in \set{1,\dots,S}$.

\vspace{2mm}

\noindent \textbf{\textit{Remark 1}}: Figs. \ref{fig:optimal-number-contending-mtds}--\ref{fig:optimal-number-ris-elements} reveal 
the \textit{optimal throughput} and \textit{optimal EE} in terms of the number of time slots, $S$.
The optimal metric values are calculated knowing that $A \leq K$, and that the denominators of eqs. \eqref{eq:throughput} and \eqref{eq:energy-efficiency} are strictly increasing functions of $S$.
Hence, the optimal throughput and EE w.r.t. $S$ can be obtained by evaluating the metrics with the number of time slots in the set $\set{S \in \integerset_+ \mid A(S) \leq K}$.

\vspace{2mm}

\begin{figure}[b]
\centering
\subfigure[Throughput]{
\includegraphics[width=.22\textwidth]{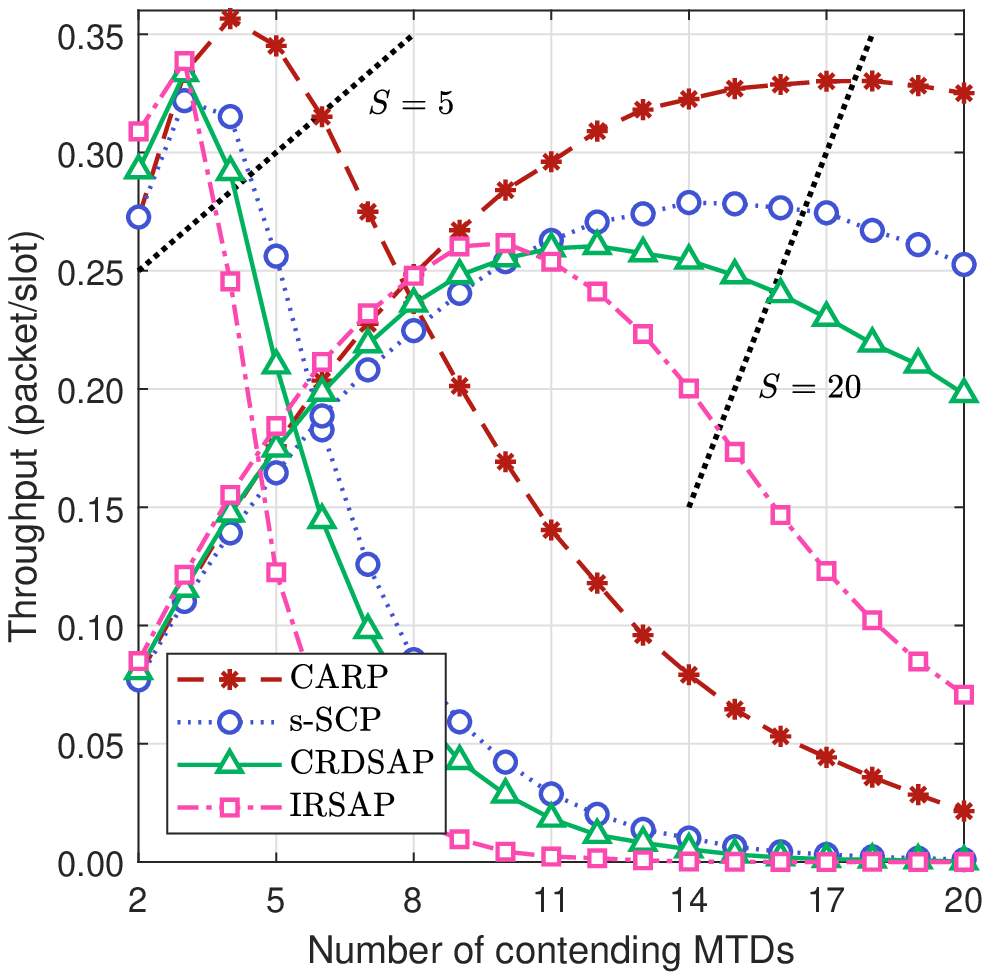}
}
\subfigure[Energy efficiency]{
\includegraphics[width=.22\textwidth]{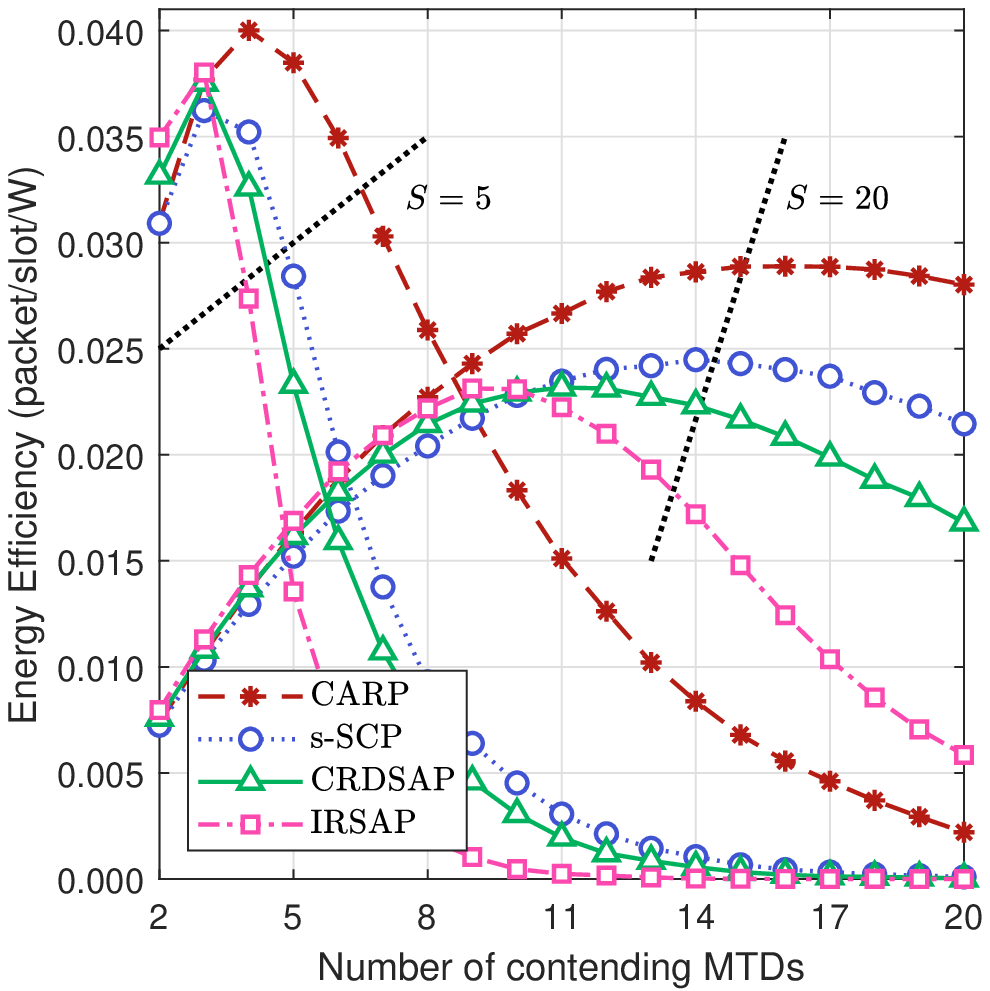}
}
\caption{a) $G\times K$ and b) $\eta_{\textsc{ee}}\times K$ for two different numbers of time slots, $S \in \set{5, 20}$. $\rho_{\textsc{mtd}} = 10$ mW, $N = 100$.}
\label{fig:number-contending-mtds}
\end{figure}

Fig. \ref{fig:number-contending-mtds} presents the throughput and EE \textit{vs.} the number of contending MTDs, $K$. Two values of number of time slots, $S$, are considered. CARP outperforms significantly the other policies both in throughput and EE for higher numbers of contending MTDs, being suitable for crowded networks. In the sequence, $s$-SCP presents an advantage over CRDSAP and IRSAP for $S = 20$ and $K > 11$. It is worth mentioning that CARP and $s$-SCP present slow throughput and EE decaying when compared to CRDSP and IRSAP. On the matter of the policies that do not require the DL training phase (CRDSAP and IRSAP), CRDSAP performance degrades with $K$ later than IRSAP.

\begin{figure}[t]
\centering
\subfigure[Throughput]{
\includegraphics[width=.22\textwidth]{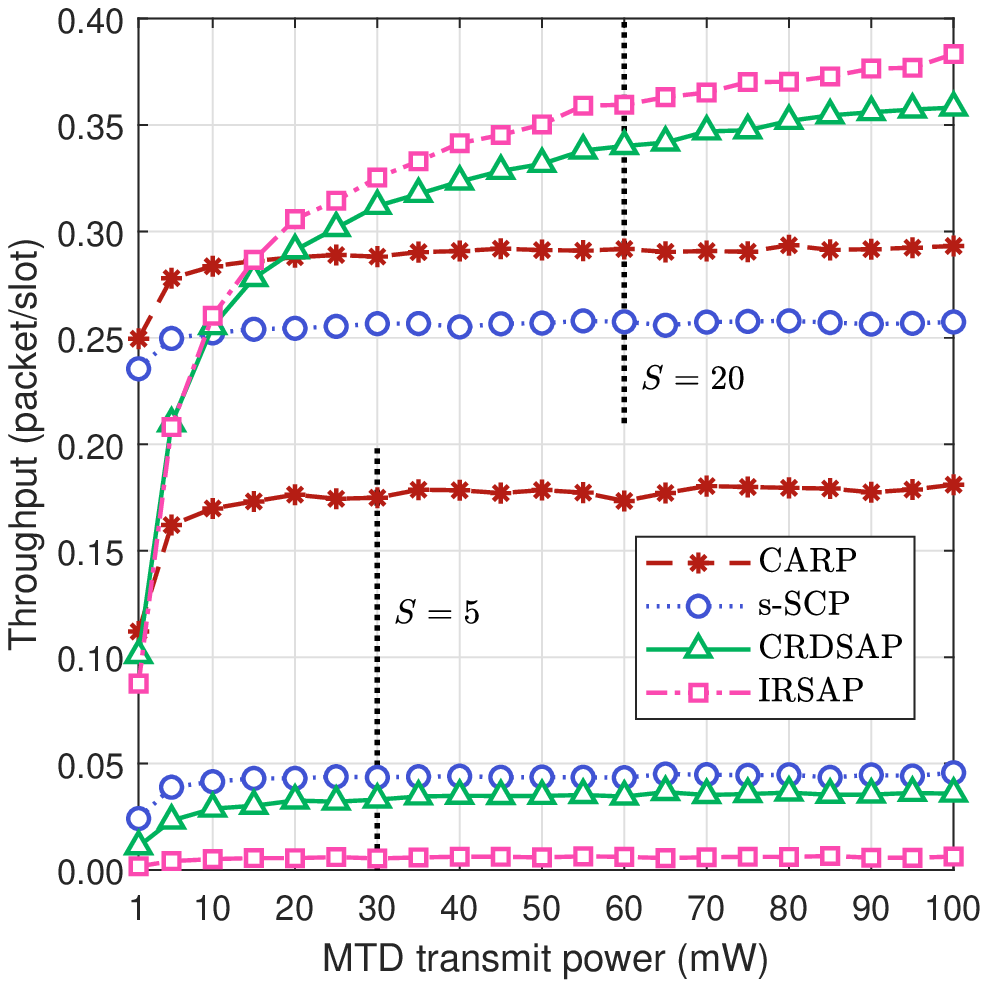}
\label{fig:mtd-transmit-power-throughput}
}
\subfigure[Energy efficiency]{
\includegraphics[width=.22\textwidth]{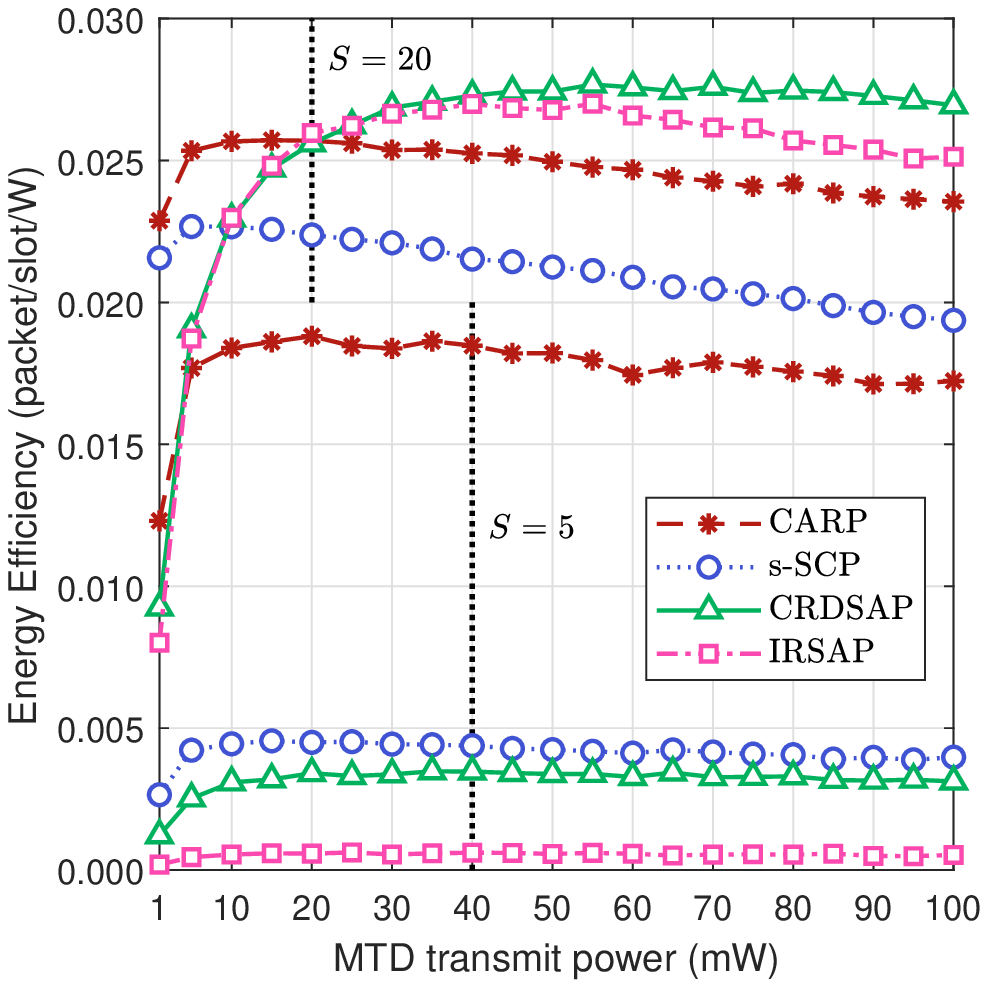}
}
\caption{a) $G \times \rho_{\textsc{mtd}}$ and b) $\eta_{\textsc{ee}} \times \rho_{\textsc{mtd}}$ for two different numbers of time slots, $S \in \set{5, 20}$. $K = 10$, $N = 100$.}
\label{fig:mtd-transmit-power}
\end{figure}

Fig. \ref{fig:mtd-transmit-power} presents the throughput and EE \textit{vs.} the MTD transmit power, $\rho_{\textsc{mtd}}$. For $S = 5$, the CARP outperforms significantly all the other policies both in throughput and EE. Still, $s$-SCP attains performance close to CRDSAP, while IRSAP achieves poor throughput and EE values.
Differently, for $S = 20$, CARP and $s$-SCP outperforms the other policies for low values of transmit power ($\rho_{\textsc{mtd}} < 20$ mW). Finally, in the region $S = 20$, $\rho_{\textsc{mtd}} > 20$ mW, IRSAP attains the best throughput, while CRDSAP attains the best EE due to the reduced number of transmitted packet replicas.

\begin{figure}[b]
\centering
\subfigure[Throughput]{
\includegraphics[width=.22\textwidth]{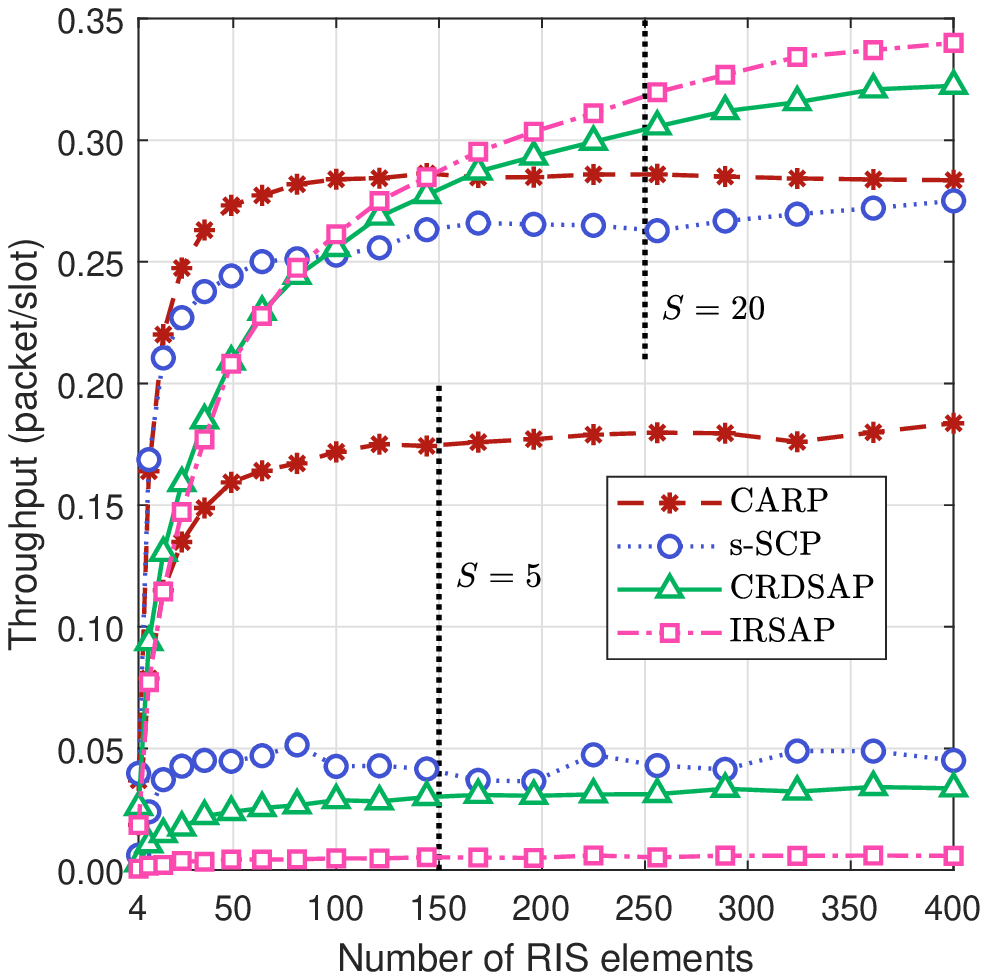}
}
\subfigure[Energy efficiency]{
\includegraphics[width=.22\textwidth]{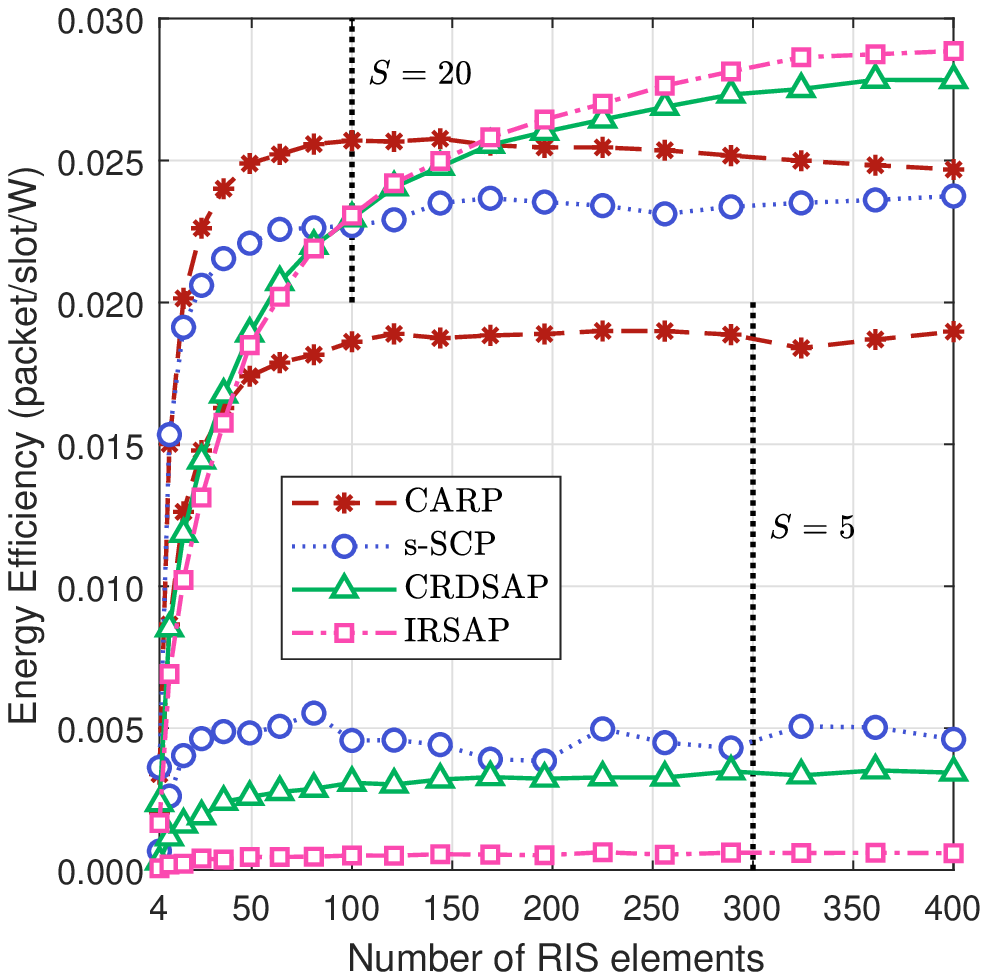}
}
\caption{a) $G\times N$ and b) $\eta_{\textsc{ee}}\times N$, for two different values of number of time slots, $S \in \set{5, 20}$. $K = 10$, $\rho_{\textsc{mtd}} = 10$ mW.}
\label{fig:number-ris-elements}
\end{figure}

Fig. \ref{fig:number-ris-elements} shows the throughput and EE \textit{vs.} the number of RIS elements, $N$. It is worth mentioning that we consider equal numbers of RIS elements in each direction, \textit{i.e.}, $N_x = N_z$.
In this result, for $S = 5$, again CARP attains remarkable performance both in throughput and EE, significantly outperforming the other policies. Also, for $S = 20$, $N < 150$, respectively CARP and $s$-SCP are the best policies. Conversely, for $S = 20$ and $N > 150$, the IRSAP becomes the best policy both in terms of throughput and EE.

% Optimal Throughput and EE w.r.t. Number of Time Slots
\subsection{Optimal Throughput and EE w.r.t. Number of Time Slots}

\begin{figure}[t]
\centering
\subfigure[Throughput]{
\includegraphics[width=.22\textwidth]{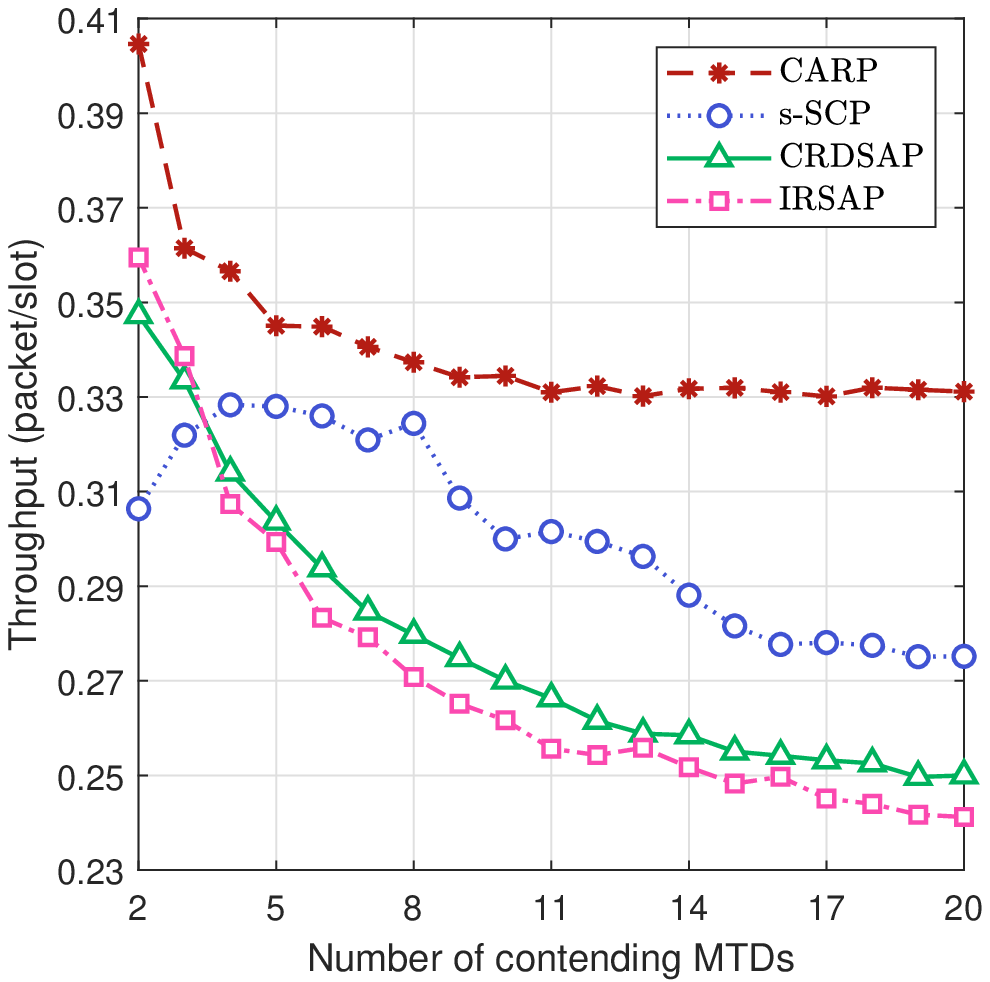}
}
\subfigure[Energy efficiency]{
\includegraphics[width=.22\textwidth]{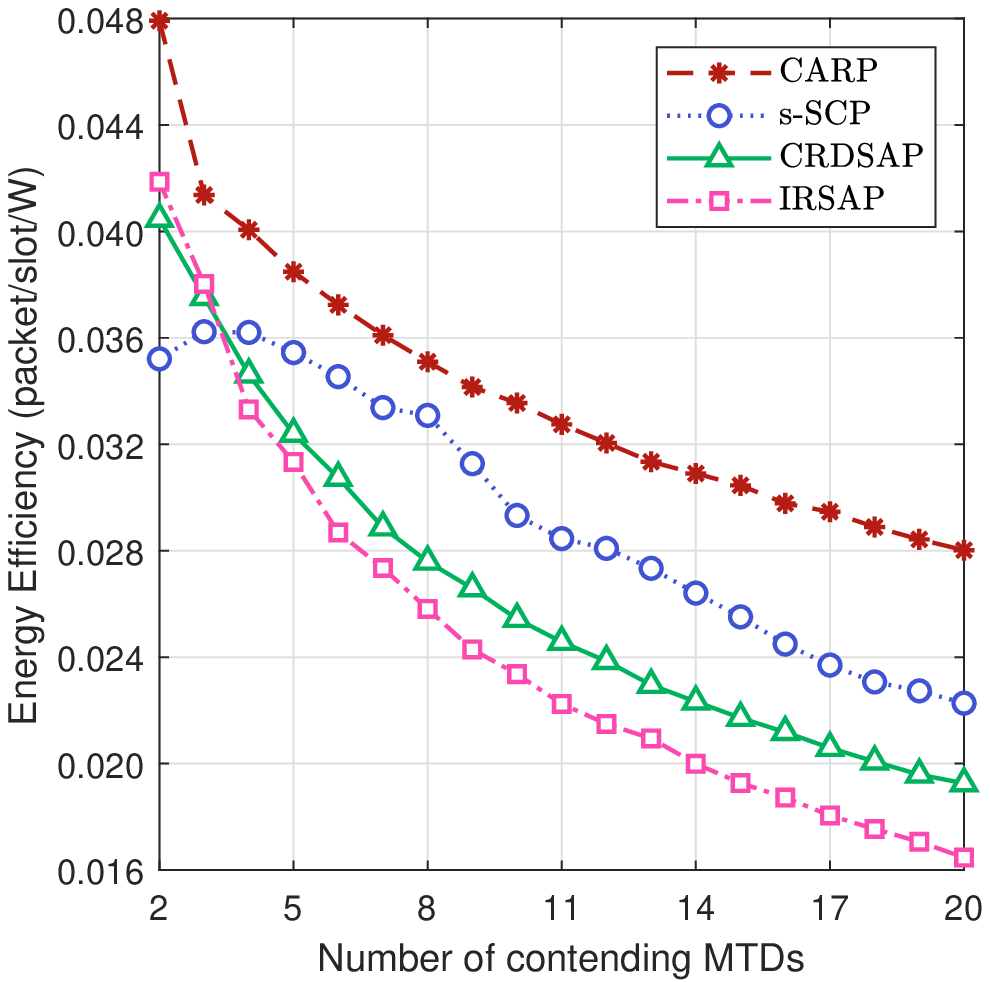}
}
\caption{a) Optimal $G$ and b) optimal $\eta_{\textsc{ee}}$ w.r.t. $S$ \textit{vs.} the number of contending MTDs, $K$. $\rho_{\textsc{mtd}} = 10$ mW, $N = 100$.}
\label{fig:optimal-number-contending-mtds}
\end{figure}

\begin{figure}[b]
\centering
\subfigure[Throughput]{
\includegraphics[width=.22\textwidth]{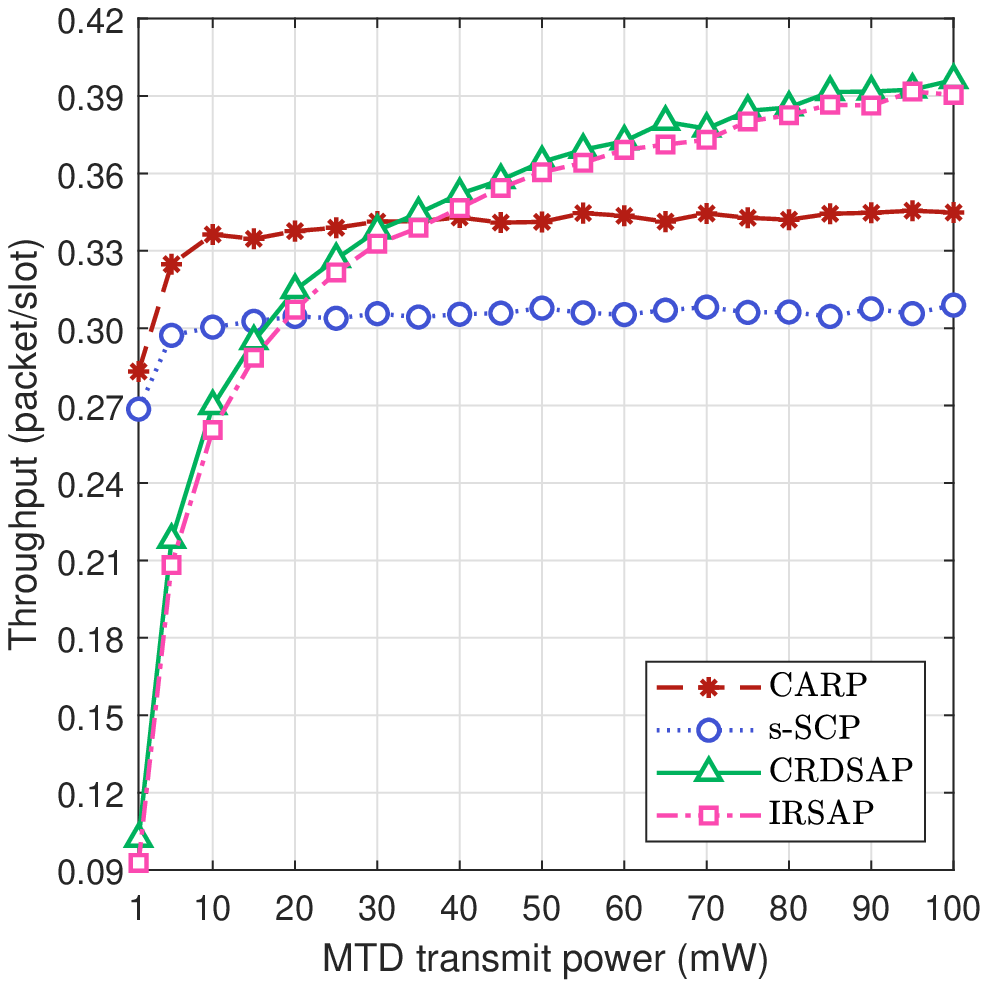}
}
\subfigure[Energy efficiency]{
\includegraphics[width=.22\textwidth]{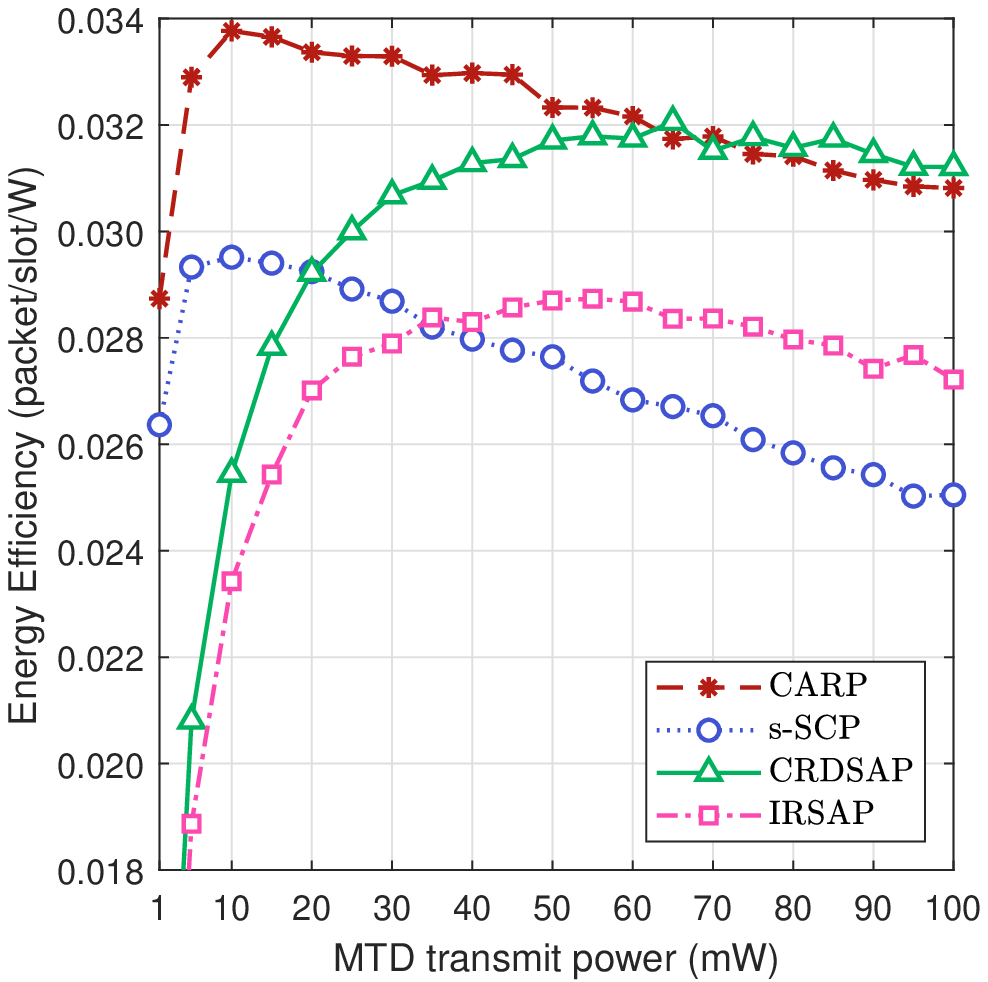}
}
\caption{a) Optimal $G$ and b) optimal $\eta_{\textsc{ee}}$ w.r.t. $S$ \textit{vs.} the MTD transmit power, $\rho_{\textsc{mtd}}$. $K = 10$, $N = 100$.}
\label{fig:optimal-mtd-transmit-power}
\end{figure}

\begin{figure}[t]
\centering
\subfigure[Throughput]{
\includegraphics[width=.22\textwidth]{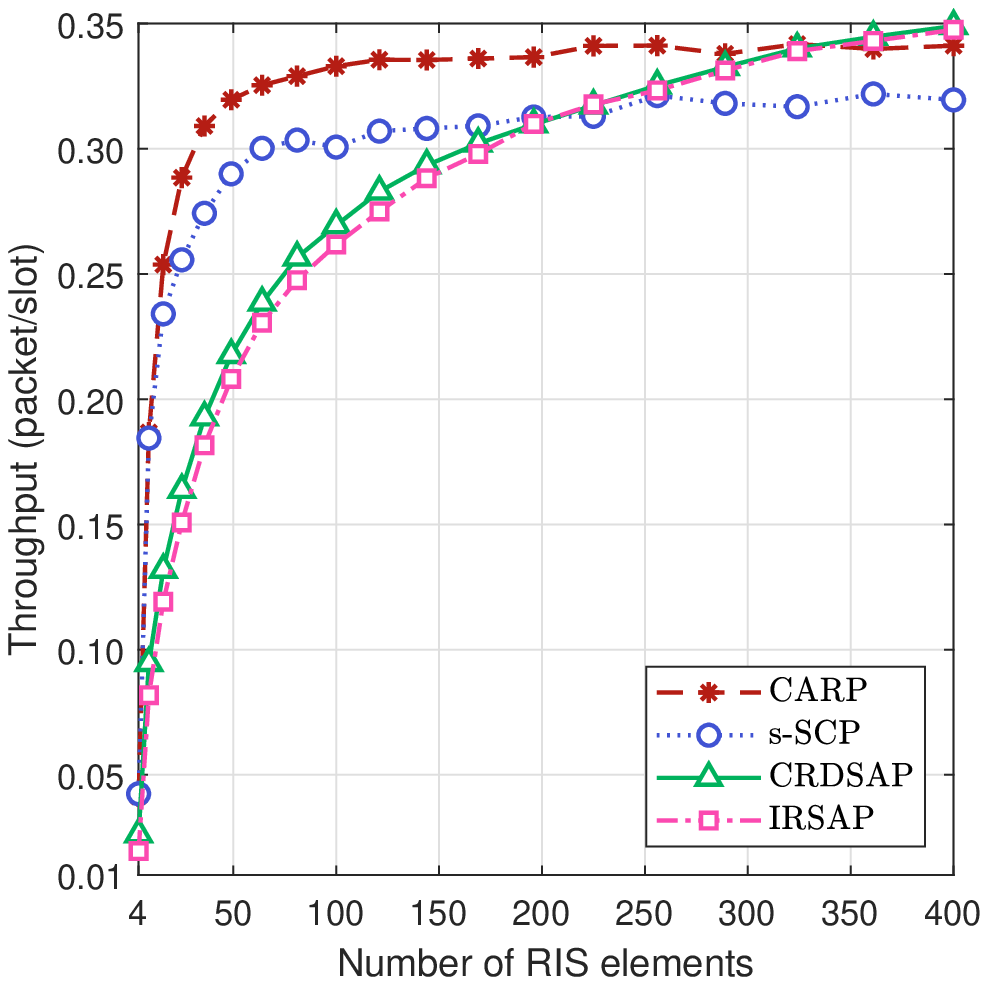}
}
\subfigure[Energy efficiency]{
\includegraphics[width=.22\textwidth]{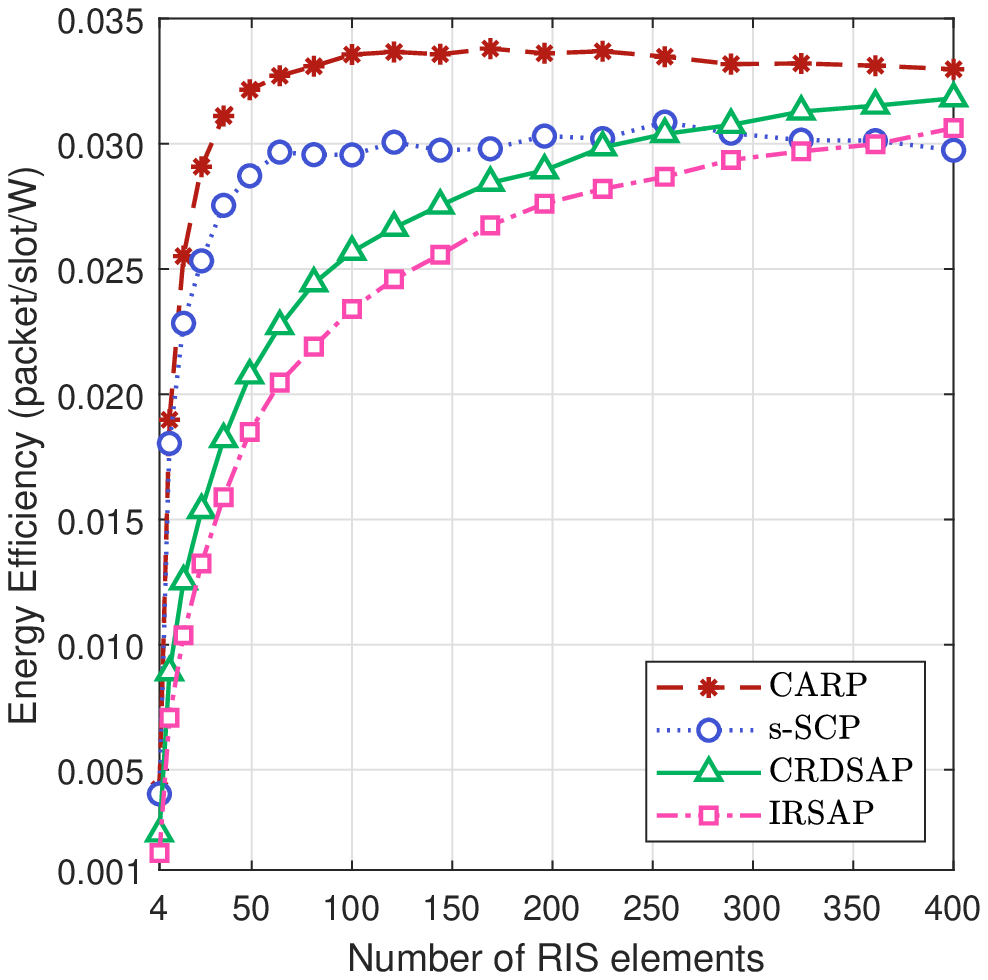}
}
\caption{a) Optimal $G$ and b) optimal $\eta_{\textsc{ee}}$ w.r.t. $S$ \textit{vs.} the number of RIS elements, $N$. $K = 10$, $\rho_{\textsc{mtd}} = 10$ mW.}
\label{fig:optimal-number-ris-elements}
\end{figure}

In the next three figures, we present the \textit{optimal throughput} and \textit{optimal EE} w.r.t. the number of time slots, $S$, considering increasing the a) number of contending MTDs (Fig. \ref{fig:optimal-number-contending-mtds}), b) MTD transmit power (Fig. \ref{fig:optimal-mtd-transmit-power}), and c) number of RIS elements (Fig. \ref{fig:optimal-number-ris-elements}).
The optimal throughput and EE w.r.t. the number of time slots, $S$, \textit{vs.} the number of contending MTDs, $K$, is revealed in Fig. \ref{fig:optimal-number-contending-mtds}. One can see that the attained throughput for all four policies decreases with $K$ until reaching a floor. Conversely, the EE is monotonically decreasing with $K$, since the power consumption is proportional to the number of contending MTDs.
Looking at the throughput floor attained by CARP, one can notice its outstanding advantage to operate in crowded networks over the other policies.

Fig. \ref{fig:optimal-mtd-transmit-power} presents the optimal throughput and EE w.r.t. the number of time slots, $S$, \textit{vs.} the MTD transmit power, $\rho_{\textsc{mtd}}$. In this result, the throughput analysis is similar to that of Fig. \ref{fig:mtd-transmit-power-throughput}, except that the IRSAP performance is near the same as CRDSAP.
Differently, under the EE perspective, the CARP policy attains remarkable results for low values of transmit power ($\rho_{\textsc{mtd}} < 60$ mW), being a promising access policy for \textit{very low-power} IoT networks. Still on EE, one can see that, for higher transmit power values, the IRSAP performance is significantly degraded when compared with CRDSAP. This is due to the fact that IRSAP allows the transmission of up to $S$ packet replicas by each contending MTD, increasing significantly the number of transmitted replicas during the UL access phase.

Fig. \ref{fig:optimal-number-ris-elements} reveals the optimal throughput and the optimal EE w.r.t. the number of time slots, $S$, \textit{vs.} the number of RIS elements, $N$. Again, the CARP policy attains the best EE when compared with the other policies. It is worth mentioning that, after the maximum value is reached, the EE slowly decreases with $N$, since the power required by each RIS element is quite low. This demonstrates the advantage of RIS in providing reasonable network performance at the cost of an extremely-low power consumption when compared with conventional network solutions that use active elements.

Analyzing Figs. \ref{fig:optimal-number-contending-mtds}--\ref{fig:optimal-number-ris-elements}, one can conclude that the RIS-aided communication is an energy-efficient technology suitable for green IoT networks, providing high throughput gains with a marginal impact on the EE, even for surfaces with large numbers of reflecting elements.
In addition, the \textit{access policies} that exploit information of the channel qualities contribute significantly with the reduction of the power consumption at the MTDs-side. This is paramount in the proposed IoT network setup, since the MTDs are powered by limited-power sources such as batteries, while the AP and RIS may be connected to the power grid.

% Conclusions
\section{Conclusions}\label{sec:conclusions}
In this paper, we develop a power consumption model, the throughput formula, and an expression for the overall EE for RIS-aided IoT networks that adopt a two time-scale RA protocol. Moreover, we adapt schemes from \cite{croisfelt2022}, \cite{casini2007}, and \cite{liva2011} for respectively developing the access policies $s$-SCP, CRDSAP, and IRSAP. The numerical results on the RA protocol performance reveal that the access policies that exploit information of the channel qualities are suitable for green IoT networks, simultaneously attaining competitive EE and throughput with low power consumption at the MTDs-side.

% References
\bibliographystyle{IEEEtran}
\bibliography{references}

\end{document}